# Autonomous Radiotherapy Treatment Planning Using DOLA: A Privacy-Preserving, LLM-Based Optimization Agent

Humza Nusrat[1,2], Bing Luo[1], Ryan Hall[1], Joshua Kim[1], Hassan Bagher-Ebadian[1,2], Anthony Doemer[1], Benjamin Movsas[1,2], Kundan Thind[1,2]

(1) Department of Radiation Oncology, Henry Ford Health, Detroit, USA
(2) College of Human Medicine, Michigan State University, East Lansing, USA

## 1. Abstract

Radiotherapy treatment planning is a complex and time-intensive process, often impacted by inter-planner variability and subjective decision-making. To address these challenges, we introduce **Dose Optimization Language Agent (DOLA)**, an autonomous large language model (LLM)-based agent designed for optimizing radiotherapy treatment plans while rigorously protecting patient privacy. DOLA integrates the LLaMa3.1 LLM directly with a commercial treatment planning system, utilizing chain-of-thought prompting, retrieval-augmented generation (RAG), and reinforcement learning (RL). Operating entirely within secure local infrastructure, this agent eliminates external data sharing. We evaluated DOLA using a retrospective cohort of 18 prostate cancer patients prescribed 60 Gy in 20 fractions, comparing model sizes (8 billion vs. 70 billion parameters) and optimization strategies (No-RAG, RAG, and RAG+RL) over 10 planning iterations. The 70B model demonstrated significantly improved performance, achieving approximately 16.4% (±4.5%) higher final scores than the 8B model. The RAG approach outperformed the No-RAG baseline by 19.8% (±2.2%), and incorporating RL accelerated convergence, highlighting the synergy of retrieval-based memory and reinforcement learning. Optimal temperature hyperparameter analysis identified 0.4 as providing the best balance between exploration and exploitation. This proof of concept study represents the first successful deployment of locally hosted LLM agents for autonomous optimization of treatment plans within a commercial radiotherapy planning system. By extending human-machine interaction through interpretable natural language reasoning, DOLA offers a scalable and privacy-conscious framework, with significant potential for clinical implementation and workflow improvement.

## 2. Introduction

Radiotherapy remains a cornerstone in the treatment of prostate cancer, offering a non-invasive approach to eradicate malignant cells while preserving healthy surrounding tissues. Advanced techniques such as intensity-modulated radiation therapy (IMRT) and volumetric modulated arc therapy (VMAT) have significantly enhanced dose delivery precision, allowing clinicians to conform high doses to the planning target volume (PTV) while minimizing exposure to critical organs at risk (OARs). However, these technical capabilities have introduced new complexities in the planning process, creating a bottleneck in radiotherapy workflows.

The creation of optimal treatment plans requires meticulous balancing of competing clinical objectives—an inherently complex optimization problem with significant clinical consequences. Several studies have demonstrated that manual planning is both time-intensive, often requiring several hours per patient, and subject to substantial inter-planner variability.[1–3] This variability stems from differences in planner expertise and subjective decision-making, potentially leading to inconsistent plan quality across patients and institutions. Suboptimal plans may either under-dose the tumor, increasing recurrence risk, or over-dose OARs, leading to adverse effects such as urinary or rectal complications in prostate cancer treatment.

Recent years have seen various artificial intelligence (AI) approaches emerge to address these challenges. Convolutional neural networks (CNNs) have been employed for dose prediction, leveraging imaging data to estimate optimal dose distributions.[4-6] Knowledge-based planning systems, which utilize historical treatment plans to suggest parameters for new cases, have shown promise in streamlining the planning process. Despite these advancements, existing AI solutions face several critical limitations. First, they typically address isolated aspects of the planning process rather than providing comprehensive decision-making across the entire planning workflow. Second, they predominantly rely on centralized data processing, raising significant privacy concerns in healthcare environments governed by regulations such as the Health Insurance Portability and Accountability Act (HIPAA[7]) and the General Data Protection Regulation (GDPR[8]). These privacy constraints have created a fundamental tension between leveraging data-driven AI techniques and protecting sensitive patient information, limiting clinical adoption.

While inter-planner variability in radiotherapy planning is often viewed negatively due to potential inconsistencies in plan quality, variability stemming from experienced clinical judgment and patient-specific considerations can be beneficial. Experienced planners frequently adjust plans to patient-specific characteristics, such as anatomical anomalies or unique clinical factors like organ dysfunction, that standardized or rigid automated models may overlook.[2,3] Indeed, clinical evidence supports the notion that treatment plans customized to individual patient needs can lead to improved clinical outcomes and reduced morbidity.[9–11]

Conversely, a critical limitation of many current AI-driven planning solutions, particularly those relying on deep learning or knowledge-based planning, is their inflexible and black-box nature. While these systems achieve consistency, they often fail to explain their optimization processes clearly, making it difficult for clinicians to validate or trust their recommendations.[12] Moreover, automated plans that seem optimal according to numerical metrics sometimes prove practically undeliverable due to machine limitations, such as excessively complex multileaf collimators (MLC) movements or unrealistic dose gradients.[13,14] Such "optimal yet impractical" AI-generated plans can introduce significant challenges in clinical translation, highlighting a crucial gap between theoretical optimization and practical clinical delivery.

These limitations underline the need for AI planning systems capable of integrating patient-specific clinical reasoning and offering transparent decision-making processes. Large Language Model (LLM)-based agents, leveraging natural language and explicit reasoning frameworks, present a promising alternative. Unlike traditional black-box AI, LLM agents can inherently communicate their decision rationale through natural language, potentially bridging the gap between automated efficiency and human interpretability.

Large language models (LLMs) based on transformer architectures[15] like GPT-4[16] offer a promising alternative approach to radiotherapy planning automation. Unlike previous AI methods, LLMs excel at complex reasoning tasks that integrate diverse types of information—a capability particularly suited to radiotherapy planning.[17] These models can interpret clinical guidelines, reason through multi-step processes, and handle unstructured data, allowing them to navigate the nuanced trade-offs inherent in balancing tumor coverage with OAR sparing. Furthermore, their deployment within local computational infrastructure can address privacy concerns by eliminating the need for external data sharing. Agentic AI systems are autonomous computational entities that act within their environment using tools to make decisions that guide actions towards specific goals, all with minimal human intervention.[18,19] LLM agents represent a specialized implementation of this paradigm, where large language models serve as the cognitive core, coordinating multiple capabilities including memory management, planning, reasoning, and tool utilization to solve complex tasks. Unlike conventional AI systems that address isolated components of a workflow, these agents can orchestrate end-to-end processes by maintaining contextual awareness, formulating strategies based on domain knowledge, and iteratively refining their approaches based on feedback—capabilities particularly suited to the multifaceted optimization challenges in radiotherapy planning.

In this study, we introduce the Dose Optimization Language Agent (**DOLA**), a novel AI agent that leverages LLMs to autonomously optimize radiotherapy treatment plans while maintaining strict patient data privacy. Our approach integrates three key technical innovations to enhance the LLM's performance in this domain. First, chain-of-thought prompting guides the LLM through structured iterative reasoning for progressive plan refinement. This enables the model to systematically address planning optimization parameters in a manner similar to human experts, considering interdependencies between decisions. Second, retrieval-augmented generation (RAG) allows the LLM to access and learn from prior planning attempts stored locally, improving its ability to adapt to patient-specific challenges and refine strategies over successive iterations. Third, reinforcement learning (RL) incorporates a reward function that guides the trade-offs between target coverage and organ at risk sparing. This reward structure ensures generated plans adhere to established clinical standards while providing a consistent optimization framework.

## 3. Methods

Radiotherapy treatment planning, particularly for advanced techniques such as IMRT and VMAT, traditionally requires expert planners to manually adjust parameters to balance competing clinical objectives. This process is inherently time-consuming and prone to errors as outcomes often depend on the planner's expertise and subjective decision-making. To address these limitations, we developed DOLA leveraging LLMs to automate and refine the radiotherapy planning process.

### 3.1. System Architecture and Design Principles

Our framework integrates the LLaMa3.1[20] LLMs (Meta AI, Menlo Park, CA) with the Eclipse treatment planning system (TPS; V16.1) (Varian Medical Systems, Palo Alto, CA). DOLA consists of three interconnected components working in concert to achieve clinically acceptable treatment plans. The working memory module maintains the current state of the treatment plan and, when configured, retrieves historical planning data to inform decision-making. The LLaMa3.1 LLM serves as the decision-making engine, analyzing the plan state and proposing adjustments to optimization parameters based on clinical goals. Finally, the TPS interface tools enable the LLM to interact directly with the Eclipse TPS by modifying the priority numbers of optimization objectives. Privacy protection was a foundational design principle of our framework. To ensure a patient privacy-centric approach, we deployed the LLM locally within the hospital's secure computational infrastructure. This approach eliminates the need for external data sharing, minimizing the risk of data breaches while maintaining the performance benefits of advanced AI techniques. By keeping all patient data within a controlled environment, the framework adheres to stringent ethical and legal standards, facilitating its potential for clinical adoption.

The overall system architecture is presented in Figure 1, comprising two main components: the Model Service and the Optimization Agent. At the core of the Model Service, the LLM engine manages the loading of models and performs inference tasks using in-house graphical processing units (GPUs). Currently, the virtualized large language model (VLLM) inference engine is utilized due to its superior performance, achieved through optimized configurations such as employing tensor parallelism instead of pipeline parallelism to enhance flexibility and GPU memory utilization. This service encapsulates LLM functionality behind a user-friendly application programming interface (API), allowing seamless integration with existing in-house clinical applications without necessitating additional hardware or software deployment. Furthermore, this architecture supports model distribution across multiple computational nodes, facilitating scalability.

The Optimization Agent consists of several loosely coupled components designed for autonomous operation. Central to this agent is the client component, responsible for synthesizing and formatting prompts for direct communication with the LLM. Enhanced capabilities, such as context-aware session management, are provided through an extension library. An action policy module sits atop the client, guiding decisions for each optimization iteration by constructing prompts, querying the LLM, interpreting responses, and generating actionable instructions. These instructions are passed through an action adapter, enabling interactions with the Eclipse treatment planning system via a dedicated interface. The information provided to the LLM during optimization strictly includes current optimization objectives, actual dose metrics, clinical goals, and plan quality scores, explicitly excluding patient identifiers or other sensitive data to maintain patient privacy and data security.

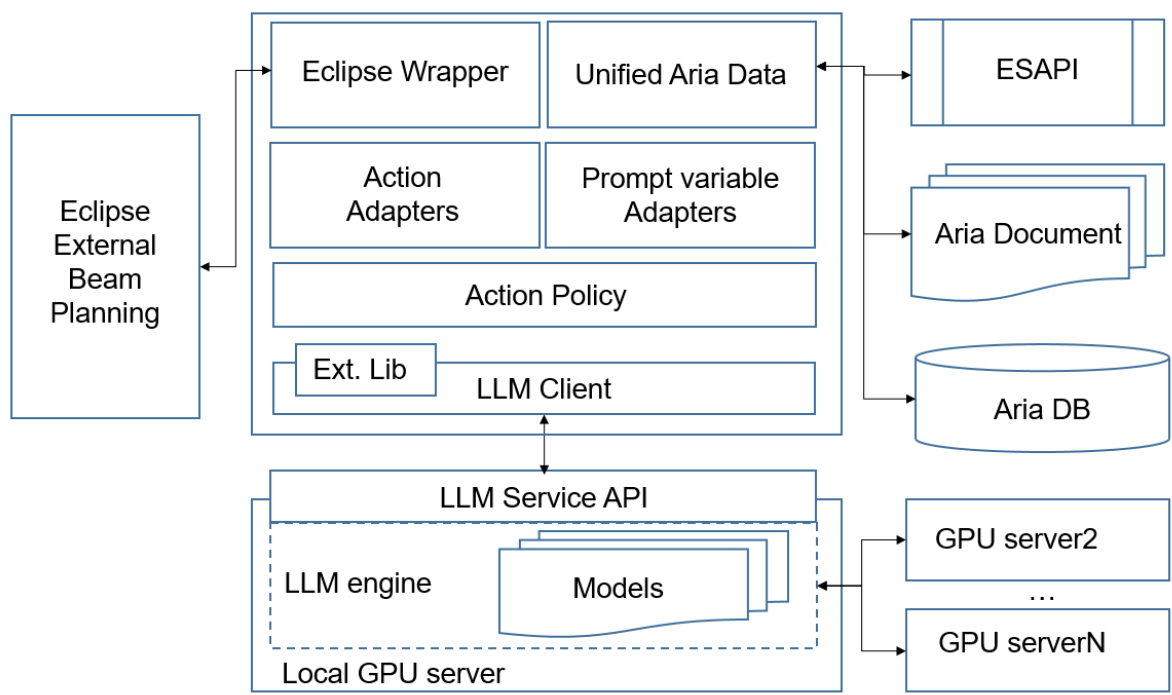

**Figure 1.** System architecture illustrating the integration and workflow of the Dose Optimization Language Agent (DOLA). The framework comprises two main components: (1) the Model Service, responsible for loading, managing, and performing inference with the LLaMa3.1 large language model (LLM) deployed locally using optimized GPU configurations, and (2) the Optimization Agent, which includes modules for prompt synthesis, action

determination, and interaction with the Eclipse treatment planning system via defined software interfaces. This architecture enables DOLA to autonomously perform iterative dose optimization while maintaining secure, privacy-preserving clinical data management.

### 3.2. Optimization Process and Parameters

For each patient, DOLA conducted a maximum of 10 iterations to generate a clinically acceptable treatment plan. This limit was established based on preliminary testing that showed convergence patterns typically stabilized within this range while maintaining computational efficiency. The iterative optimization process followed a structured sequence of steps to ensure systematic refinement of the treatment plan.

At each iteration, the LLM first assessed the current treatment plan state, including dose-volume metrics for PTVs and OARs. When RAG was enabled, the agent retrieved up to three prior planning attempts from working memory to inform decision-making. Based on this comprehensive assessment, the LLM determined appropriate modifications to the priority numbers of optimization objectives. These modifications were then implemented in the TPS, which recalculated the dose distribution according to the adjusted parameters. The resulting plan was evaluated using a scoring metric derived from the PROFIT[21] clinical trial protocol, and prioritized PTV coverage while incorporating penalties for violations of OAR dose constraints and PTV hotspot limits. This approach ensured alignment with established clinical standards for prostate radiotherapy.

DOLA's actions were deliberately constrained to modifying optimization priorities, allowing for a focused evaluation of its ability to navigate complex clinical trade-offs. This iterative approach enabled the agent to progressively refine the treatment plan, with each iteration informed by the outcomes of previous attempts.

### 3.3. Advanced Optimization Strategies

We implemented and evaluated two advanced strategies to enhance DOLA's planning capabilities. Retrieval-augmented generation (RAG) enabled it to access and learn from historical planning attempts stored in the working memory. By retrieving up to three prior iterations, DOLA could leverage successful strategies or avoid previously identified suboptimal paths, improving its adaptability over successive iterations.[22] This approach proved particularly valuable for complex cases where initial iterations failed to meet clinical thresholds, as it allowed the agent to incorporate lessons learned from past attempts.

In parallel, we implemented a reinforcement learning (RL) approach to guide the decision-making process (Equation 1). The reward function was carefully designed to balance competing clinical objectives with a structure that assigned a substantial reward for meeting the primary PTV coverage goal (≥95% of PTV receiving 58.5 Gy) and imposed a significant penalty if this goal was not achieved. Additionally, incremental rewards were provided for satisfying secondary clinical goals, including OAR dose constraints. The cumulative nature of the reward function across iterations incentivized the agent to prioritize PTV coverage early in the process before focusing on secondary objectives, mirroring clinical best practices where adequate tumor coverage is paramount, followed by the minimization of normal tissue exposure.

In LLMs, the temperature hyperparameter plays a critical role in modulating the exploration-exploitation trade-off during decision-making.[17,23–26] To optimize this parameter for radiotherapy planning, we conducted a systematic evaluation across temperatures ranging from 0.1 to 1.0, with increments of 0.1. For each temperature setting, DOLA performed 10 iterations of plan refinement on the patient cohort. We assessed performance using the PROFIT-based scoring metric and visualized results as a heatmap correlating temperature settings, iteration numbers, and relative plan scores (Figure 2). This approach allowed us to identify the optimal temperature value (0.4) that balanced exploration of diverse planning strategies with exploitation of effective approaches. This optimal temperature was used for all subsequent experiments in the study.

### 3.4. Patient Cohort and Experimental Design

We utilized a retrospective cohort of 18 prostate cancer patients treated between 2015 and 2023 at Henry Ford Health in Detroit, Michigan. All patients received a hypofractionated regimen of 60 Gy in 20 fractions, consistent with the PROFIT[21] trial protocol. The dataset included computed tomography (CT) scans, segmented structures (PTV, rectum, and bladder), and predefined clinical dose constraints. The use of retrospective patient data was conducted under institutional review board (IRB) approval, ensuring adherence to ethical standards and protecting patient confidentiality.

We evaluated DOLA across multiple configurations to assess the impact of model size and advanced techniques. For model size comparison, we tested two versions of the LLaMa3.1[20] model: the 8 billion parameter (8B) version and the 70 billion parameter (70B) version. This comparison aimed to determine the influence of model capacity on planning efficiency and plan quality, with results depicted in Figure 3. Larger model sizes were hypothesized to improve the agent's ability to navigate complex optimization landscapes, though computational efficiency was also considered.

For optimization strategy comparison, we evaluated three distinct configurations. In the baseline configuration (No-RAG), the LLM operated without retrieving historical planning data, relying solely on its internal decision-making capabilities. In the RAG configuration, DOLA was augmented with retrieval capabilities to incorporate insights from prior iterations, enhancing its adaptability. Finally, in the RAG+RL configuration, both retrieval and reinforcement learning were implemented to guide the optimization process, combining the benefits of historical learning and clinical alignment.

For each configuration, we conducted 10 iterations of plan optimization for all 18 patients. Plan quality was quantified using the PROFIT-based scoring metric (Equation 2), with both final plan scores and iteration-by-iteration progression analyzed to evaluate convergence patterns and overall performance. These analyses provided insights into the effectiveness of RAG and RL in accelerating the optimization process and improving overall plan quality, paving the way for future refinements and clinical validation.

$$R = 50 \times I_{PTV} + \sum_{j=2}^{M} I_j - 50 \times (1 - I_{PTV})$$

$$Cumulative\ Reward = \sum_{i=0}^{10} R$$

- M: Total number of clinical goals (1 for PTV coverage, the rest for OARs)
- $I_{PTV}$: Indicator variable for PTV coverage, equal to 1 if primary coverage is greater than or equal to 95%, 0 otherwise.
- $I_J$: Indicator variable for the j-th OAR goal, equal to 1 if the goal is met, otherwise 0.
- R: Reward calculated by equation 1.

**Equation 1:** This equation describes the reward system used to evaluate the achievement of clinical goals during each iteration. The reward, R, is calculated based on whether the PTV coverage and other OAR metrics are met. The cumulative reward is the sum of R accumulated over all (ten) iterations.

$$Score = S_{PTV} + \sum_i S_{OAR}$$

$$\begin{cases} S_{PTV} = 100 \\ S_{PTV} = 100 \times \dfrac{PTV_{Coverage}}{95} \end{cases} \quad if \quad \begin{cases} PTV_{Coverage} \geq 95\% \\ PTV_{Coverage} < 95\% \end{cases}$$

$$\begin{cases} S_{OAR} = 10 \\ S_{OAR} = 10 \times \left(1 - \dfrac{Deviation}{Threshold}\right) \end{cases} \quad if \quad \begin{cases} OAR_{dose} \geq Threshold \\ OAR_{dose} < Threshold \end{cases}$$

$$where\ Deviation = OAR_{dose} - Threshold$$

**Equation 2:** Plan scoring metric used to quantitatively evaluate radiotherapy treatment plans produced by DOLA. The metric assigns a base score of 100 points for achieving the primary clinical goal of at least 95% planning target volume (PTV) coverage. If this threshold is not met, the PTV coverage score decreases proportionally relative to the achieved coverage. Additionally, penalties for exceeding dose-volume constraints for organs at risk (OARs) are calculated and subtracted from this base score. Specifically, OAR violations incur proportional deductions scaled by the degree of deviation from predefined dose thresholds, thus ensuring plans are rewarded for robust PTV coverage and penalized for compromising OAR safety.

## 4. Results

Our investigation of autonomous radiotherapy planning using LLMs yielded significant insights into performance optimization, model scaling effects, and dosimetric outcomes across multiple planning strategies. The results demonstrate that integrating advanced techniques like retrieval-augmented generation and reinforcement learning can substantially improve plan quality while maintaining computational efficiency.

### 4.1. Temperature Parameter Optimization for Decision Making

The temperature parameter, which modulates exploration-exploitation balance in the LLM's decision-making process, proved critical to optimization performance. Figure 2 presents a comprehensive heatmap visualization of relative plan scores as a function of temperature settings (0.1-1.0) and optimization iterations (0-10).

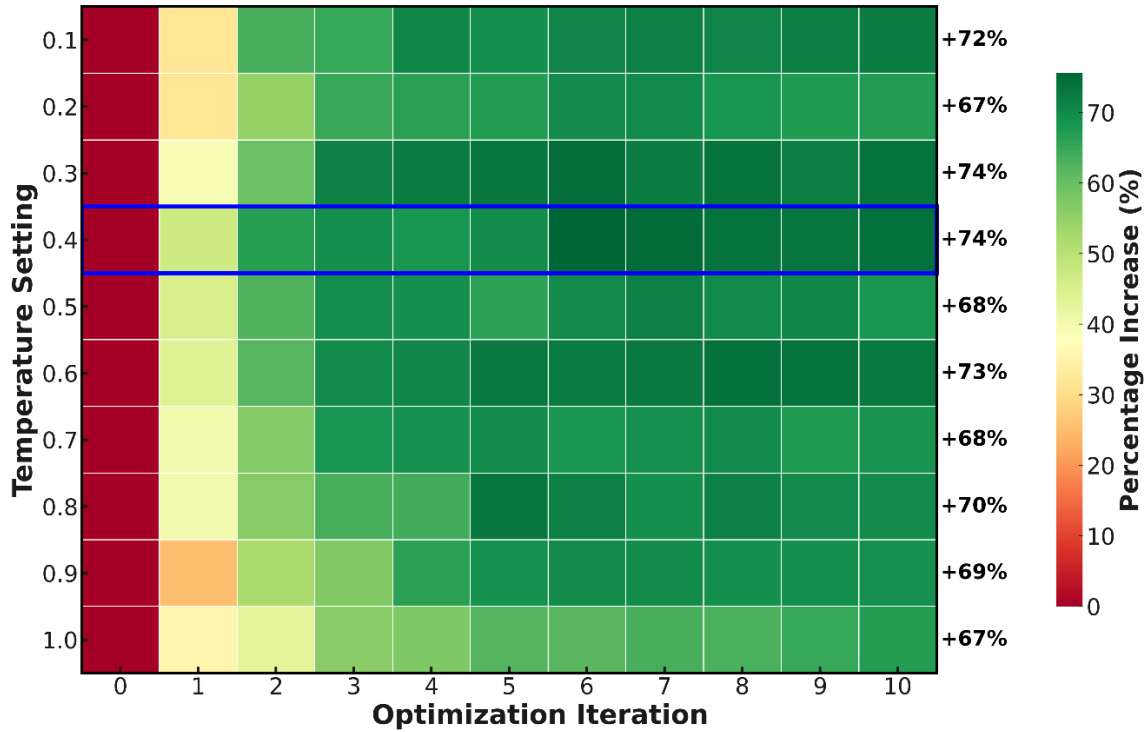

**Figure 2.** Temperature parameter optimization heat map. Within each temperature cohort, mean plan score values were normalized to the initial baseline value (iteration 0). The color scale represents relative plan score increases (from baseline) as a function of temperature settings (y-axis, 0.1-1.0) and optimization iterations (x-axis, 0-10). Lower temperatures (0.1-0.4) yield progressively higher plan scores in later iterations, with optimal performance achieved at temperature 0.4 (score increase of +74%). The blue square indicates the first temperature setting to achieve 95% PTV coverage (T=0.4 at iteration 4). Higher temperatures (0.8-1.0) demonstrate delayed or reduced improvement, reflecting suboptimal exploration-exploitation balance.

As iterations progressed, scores improved across most temperature settings, with the highest-performing region consistently appearing at lower temperatures (0.1-0.4). The optimal performance occurred at temperature 0.4 during iteration 10, achieving a maximal score increase of +74% (Figure 2). Higher temperature settings (0.8-1.0) produced inferior results, particularly in early iterations, suggesting that excessive exploration hindered convergence toward optimal solutions. This finding suggests that moderate temperature values strike an optimal balance between exploration and exploitation for radiotherapy planning tasks. Based on these results, we selected temperature 0.4 as the standard setting for all subsequent experiments, providing an ideal balance between optimization performance and clinical threshold achievement.

## 4.2. Impact of Large Language Model Scale on Planning Performance

Model scale substantially influenced planning performance, with the larger 70B parameter model demonstrating clear advantages over the smaller 8B parameter version. Figure 3 compares the relative plan scores over ten optimization iterations for both models under identical No-RAG conditions, isolating the effect of model scale.

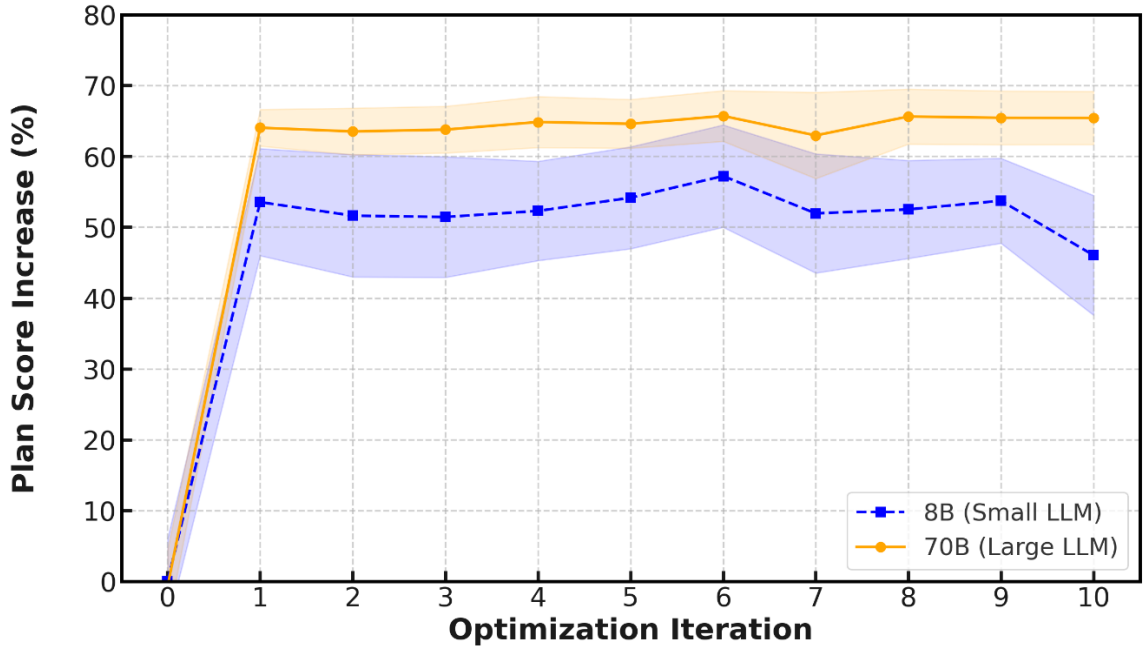

**Figure 3.** Impact of model scale on DOLA's performance. Relative plan scores (y-axis) across 10 optimization iterations (x-axis) comparing the 70B parameter model (orange line) and 8B parameter model (blue dashed line) under identical No-RAG configurations. The larger model demonstrates superior performance with faster convergence by iteration 1 and consistently higher scores maintained through iteration 10. Narrower confidence intervals for the 70B model indicate greater consistency across patients compared to the 8B model. By the final iteration, the 70B parameter model achieved approximately 16.4% (±4.5%) higher mean scores than the 8B model. Confidence intervals represent standard error of the mean calculated across all patients.

Both models began with similar baseline scores (approximately 1.0) at iteration 0. The large LLM exhibited rapid convergence, achieving a relative score increase of 60% by iteration 1 and maintaining consistent performance between 60% and 70% through iteration 10. The narrow confidence interval surrounding the 70B model's performance trajectory indicates remarkable consistency across the patient cohort. In contrast, the small LLM showed more gradual improvement, reaching only +10% by iteration 1 and requiring six iterations to approach its peak performance. The wider confidence interval for the 8B model reflects greater variability in its planning outcomes, suggesting less reliable performance across different patient anatomies. Furthermore, the small model's performance declined slightly in later iterations (8-10), highlighting potential limitations in maintaining solution stability. By iteration 10, the 70B parameter model achieved a final mean score approximately 16.4% (±4.5%) higher than the 8B parameter model. This substantial performance gap demonstrates that increased model capacity significantly enhances the ability to navigate the complex, multi-objective optimization landscape of radiotherapy planning, likely due to improved reasoning capabilities and greater contextual understanding.

### 4.3. Comparative Evaluation of Optimization Strategies

To assess the impact of our proposed advanced techniques, we compared three distinct optimization strategies: baseline (No-RAG), RAG, and the combined approach of RAG+RL. Figure 4 illustrates the progression of relative plan scores across ten iterations for each strategy using the 8B parameter model.

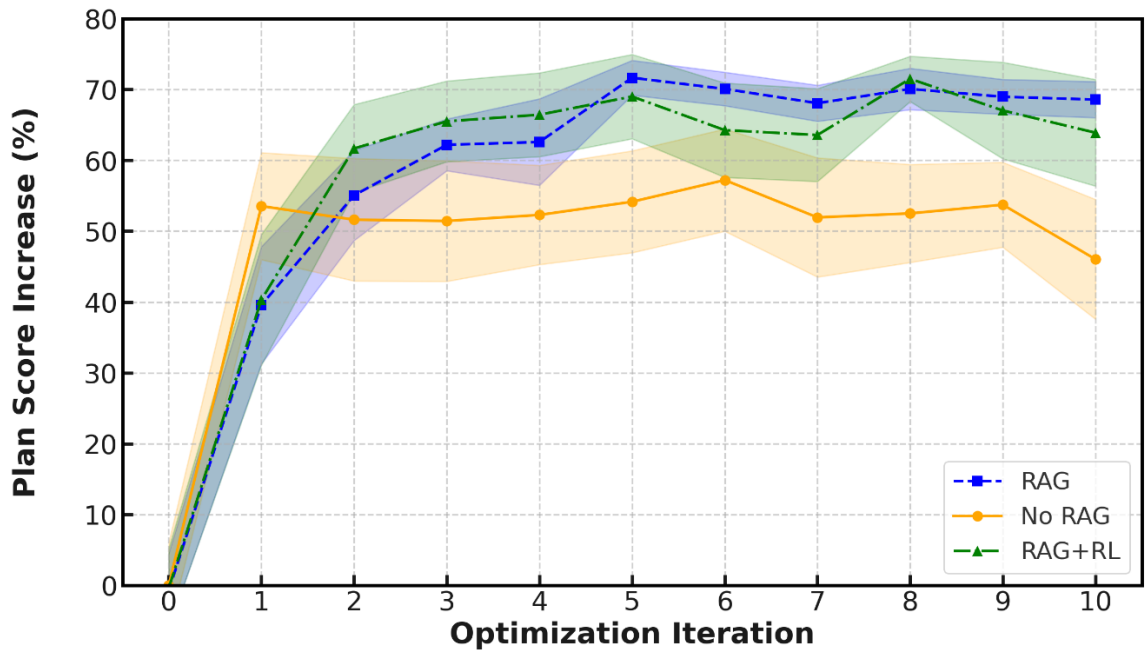

**Figure 4.** Comparative analysis of optimization strategies for DOLA. Relative plan scores (y-axis) over 10 iterations (x-axis) for three approaches: No-RAG baseline (orange), RAG (blue dashed), and RAG+RL (green dash-dot). The No-RAG configuration shows rapid initial improvement followed by plateau at approximately +50%. The RAG configuration demonstrates more gradual but sustained improvement (19.8% ±2.2% higher than No-RAG at the final iteration). The RAG+RL configuration exhibits the most dynamic trajectory, peaking at iteration 7 with faster convergence than RAG alone. Wider confidence intervals for RAG+RL

reflect greater exploration of the optimization space. All experiments conducted with the 8B parameter model at temperature 0.4.

The No-RAG configuration demonstrated rapid initial improvement followed by minimal further gains. This pattern suggests that the baseline approach quickly exhausts its optimization potential without access to historical planning data. The RAG configuration displayed a more gradual but sustained improvement trajectory representing an 19.8% (±2.2%) improvement over the No-RAG baseline. Most notably, the RAG+RL configuration exhibited a distinctive optimization pattern, combining rapid early improvements with high peak performance. This approach achieved the highest mid-process scores, peaking at iteration 7, before slightly declining. The wider confidence interval for RAG+RL indicates greater exploration of the solution space, consistent with the reinforcement learning component encouraging broader sampling of planning strategies.

These results highlight the complementary benefits of combining retrieval-based memory with reinforcement learning. RAG enhances the model's ability to learn from previous iterations, while RL accelerates the discovery of high-quality solutions through strategic exploration guided by clinically aligned rewards. The synergistic effect is particularly evident in the RAG+RL configuration's faster convergence compared to RAG alone.

### 4.4. Dosimetric Performance across Optimization Iterations

The clinical efficacy of our approach is ultimately determined by its ability to meet specific dosimetric criteria for target coverage and organ sparing. Figure 5 presents a comprehensive visualization of key dosimetric parameters across optimization iterations for the RAG+RL method, including PTV coverage and dose constraints for bladder and rectum.

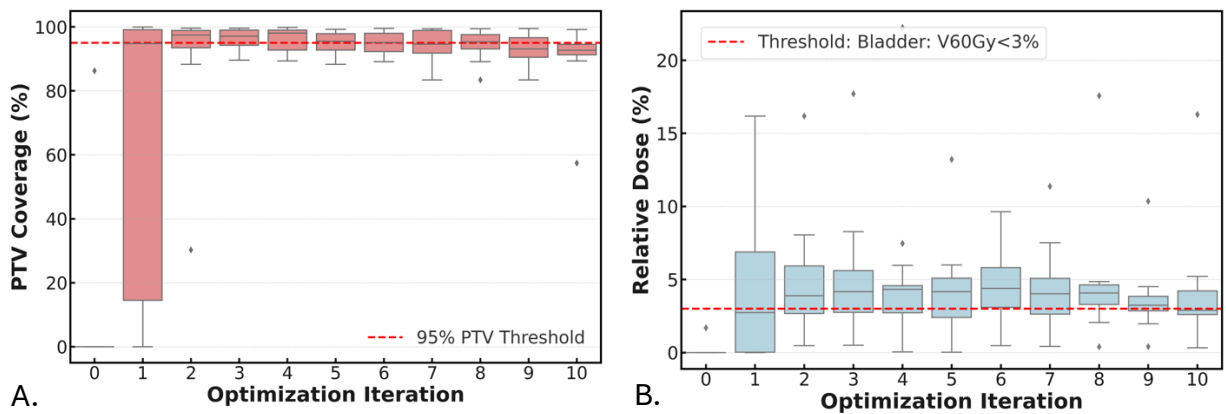

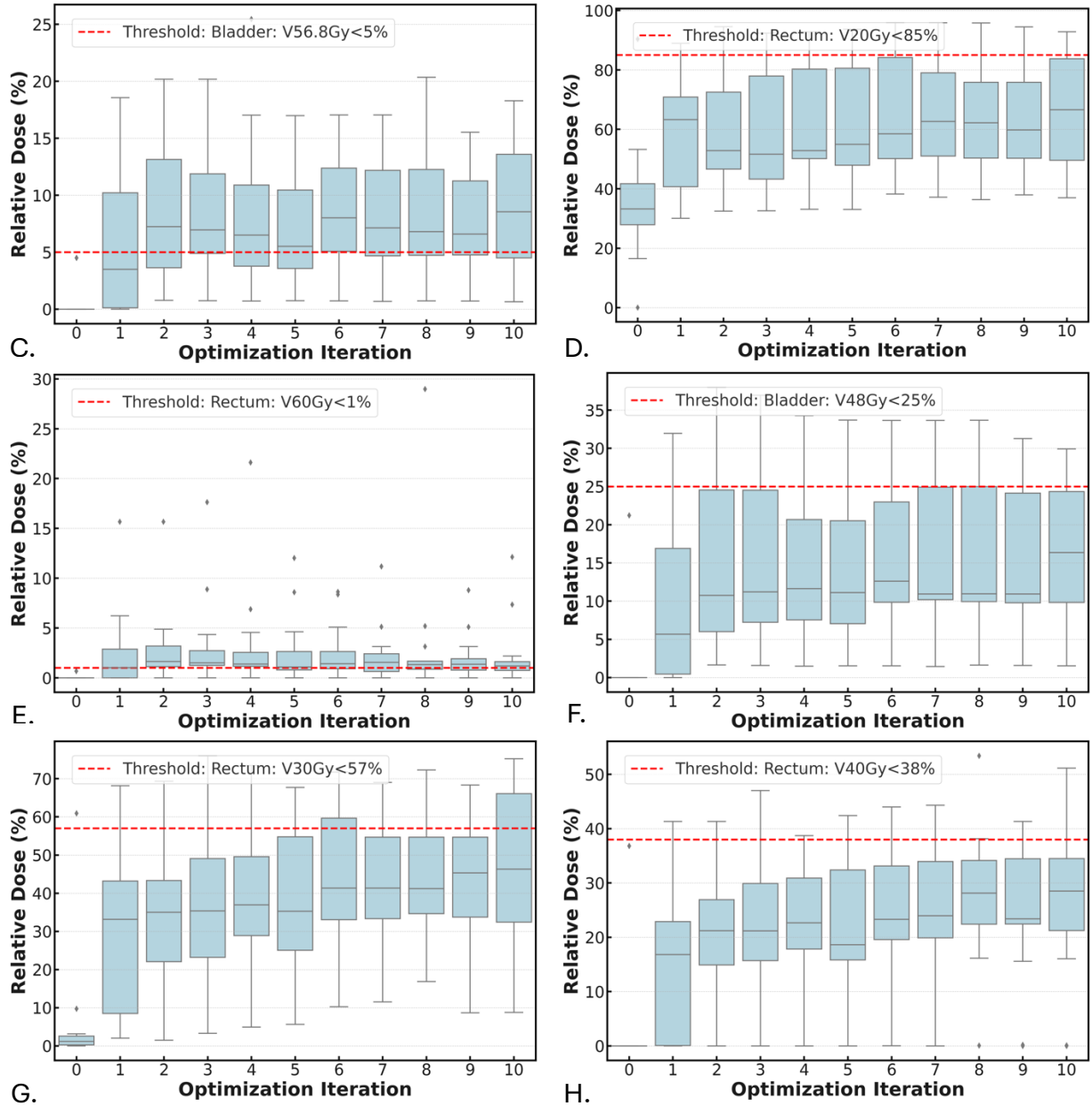

**Figure 5.** Dosimetric outcomes across optimization iterations for the RAG+RL method using the 8B model. Box-and-whisker plots showing the progression of key dose metrics over 10 iterations (0-10) for the 18 prostate cancer patient cohort: (A) PTV Coverage (%), demonstrating rapid achievement of the 95% clinical threshold by iteration 1; (B) Bladder V60Gy (<3%), showing significant reduction in high-dose bladder volume; (C) Bladder V48Gy (<25%); (D) Bladder V56.8Gy (<5%); (E) Rectum V60Gy; (F) Rectum V20Gy (<85%); (G) Rectum V30Gy (<57%); and (H) Rectum V40Gy (<38%),. Horizontal red lines represent clinical thresholds from the PROFIT trial protocol. Box boundaries indicate interquartile ranges (25th-75th percentiles), center lines represent medians, whiskers extend to 1.5 times the interquartile range, and points represent outliers.

PTV coverage (Figure 5A) demonstrated drastic improvement from a median of approximately 20% at iteration 0 to consistently above 95% from iteration 1 onward, meeting the clinical requirement of 95% PTV coverage. The narrow interquartile range (typically <10%) indicates highly consistent performance across the patient cohort. For bladder dose constraints, the V60Gy metric (Figure 5B) showed substantial reduction from a median of approximately 7.5% at iteration 0 to 2.5% by iteration 1, subsequently stabilizing below the clinical threshold of 3% through iteration 10. Similar patterns were observed for other bladder constraints, including V48Gy<25% (Figure 5C), and V56.8Gy<5% (Figure 5D), with consistent convergence toward or below clinical thresholds. While occasional constraint violations occurred in later iterations—particularly for the strictest V60Gy constraint—the majority of plans successfully met all bladder sparing objectives.

Rectal dose constraints exhibited more dynamic behavior across iterations. The V60Gy metric (Figure 5E) and the V20Gy metric (Figure 5F) showed progression from iteration 0 through iteration 6, approaching but generally remaining below their respective threshold, reflecting the system's attempt to maximize target coverage while maintaining rectal sparing. Higher-dose rectal constraints including V30Gy<57% (Figure 5G) and V40Gy<38% (Figure 5H) demonstrated similar patterns of controlled progression toward clinical limits, indicating the algorithm's ability to appropriately balance competing objectives. These dosimetric results demonstrate that DOLA effectively navigates the complex trade-offs between target coverage and normal tissue sparing. The system first prioritizes achieving adequate PTV coverage—the primary clinical goal—before progressively refining OAR sparing within acceptable limits. This prioritization aligns with clinical practice, where treatment efficacy (determined by target coverage) takes precedence over reducing side effects (determined by OAR doses), provided the latter remain within established tolerance thresholds.

Collectively, these results demonstrate that our LLM-based autonomous planning framework, particularly when enhanced with RAG and RL techniques, can efficiently generate high-quality radiotherapy plans that meet clinical requirements across a range of dosimetric parameters using a smaller model. The larger 70B parameter model offers superior performance, while the integration of retrieval-augmented generation and reinforcement learning accelerates optimization and improves final plan quality.

## 5. Discussion

The proposed **DOLA** framework, integrating RAG and RL, represents a significant advancement in addressing longstanding challenges in radiotherapy workflow automation. By combining chain-of-thought prompting, RAG, and RL within a locally deployed infrastructure, DOLA achieves not only clinically acceptable treatment plans but does so while rigorously safeguarding patient data privacy. A key strength of our approach is its privacy-centric design, ensuring that all LLM-driven data processing occurs entirely within the hospital's secure computational environment. This fully local deployment strategy eliminates external data sharing, mitigating privacy risks and ensuring strict regulatory compliance.

This work is pioneering in demonstrating that large language models can effectively manage the complex, nuanced decision-making required in radiotherapy treatment planning. Our findings illustrate how LLM-based agents, when enhanced with explicit reasoning frameworks and historical data retrieval, can significantly improve planning efficiency, consistency, and overall plan quality. The technical innovation introduced by DOLA offers a clear pathway for broader integration of advanced AI methods into clinical radiotherapy practice, balancing sophisticated automation capabilities with transparency and interpretability. Previously, exploratory work on the use of LLMs for treatment planning has been conducted[27], however, these works lacked clinical translatability given that their experiments were limited to a non-clinical TPS and did not consider patient privacy concerns.

As radiotherapy evolves toward more personalized and precise treatment paradigms, frameworks like DOLA have the potential to profoundly transform clinical workflows. The integration of explainable AI within secure, privacy-focused clinical environments provides a scalable model for future research and clinical adoption. Moving forward, our approach sets a foundation for expanded studies that should include diverse patient cohorts, real-time clinical implementation, and evaluations of long-term clinical outcomes.

## 5.1. Clinical Significance of Dosimetric Outcomes

The dosimetric results demonstrate that our LLM-based planning agent consistently achieves the primary clinical objective of adequate PTV coverage while respecting organ-at-risk constraints. The rapid improvement in plan quality from iteration 0 to iteration 1, followed by more gradual refinement in subsequent iterations, mirrors the prioritization approach used by expert human planners. This pattern—first ensuring target coverage before fine-tuning OAR sparing—aligns with established clinical protocols where tumor control takes precedence over normal tissue complications, provided the latter remain within acceptable limits.

The occasional constraint violations observed in later iterations, particularly for bladder V60Gy, highlight an important characteristic of the system: its tendency to explore the boundaries of the clinically acceptable solution space. Rather than being overly conservative, the system pushes toward an optimal balance between competing objectives, occasionally exceeding constraints in individual cases while maintaining overall clinical acceptability. This behavior resembles that of experienced human planners who understand when minor deviations from guidelines may be acceptable in service of overall plan quality.[28]

Compared to existing automated planning approaches such as knowledge-based planning[4,29,30] and deep learning-based dose prediction[5,31–35], our method offers two distinct advantages. First, it addresses the entire plan optimization process rather than isolated components, providing end-to-end optimization of planning parameters. Second, it demonstrates adaptability across patients without requiring extensive training on large, site-specific datasets, suggesting broader applicability across different treatment sites and protocols.

## 5.2. Technical Innovations and Impact

Our investigation of temperature settings reveals critical insights into LLM optimization behavior in radiotherapy planning. The superior performance of lower temperature settings (0.1-0.4) indicates that moderate constraint of the model's creativity leads to better planning outcomes. This finding aligns with radiotherapy planning being a constrained optimization problem where exploration should occur within clinically relevant boundaries rather than across the entire theoretical solution space. The optimal temperature of 0.4 represents an important balance point—providing sufficient flexibility to discover diverse planning strategies while maintaining focus on clinically viable solutions. This mirrors findings in other complex decision-making domains where controlled LLM exploration outperforms either rigid determinism or excessive randomness.[36–40]

The substantial performance improvement observed with the larger 70B parameter model (16.4% higher final scores compared to the 8B model) provides empirical evidence for scaling benefits in plan optimization decision-making tasks. Larger models likely benefit from enhanced reasoning capabilities, greater contextual understanding, and improved pattern recognition—all critical for navigating the complex trade-offs in radiotherapy planning. This observation supports the growing consensus that model scale remains important for tasks requiring sophisticated reasoning, despite efforts to distill capabilities into smaller models.[41–43] While the 8B parameter model demonstrated lower baseline performance, our results show that augmentation strategies can substantially narrow this gap—the RAG configuration improved scores significantly over the No-RAG baseline, and RAG+RL demonstrated faster convergence to near-optimal solutions. These approaches effectively compensate for the more limited intrinsic reasoning capabilities of smaller models by providing external memory access and strategic optimization guidance. Test-time compute techniques, where additional computational resources are allocated during inference rather than increasing model size, represent another promising direction for enhancing small model performance.[44–46] For example, allowing more planning iterations or implementing more sophisticated sampling strategies during inference can yield significant improvements without the memory requirements of larger models. In time-sensitive clinical environments, the faster inference speed of smaller models—often quicker by a factor of 10 than their larger counterparts—translates to meaningful workflow improvements and reduced planning delays. Healthcare institutions face varying resource constraints, and the lower memory and processing demands of compact models (8B parameters requiring approximately 16GB of memory versus 140GB for 70B models) enhance accessibility across diverse clinical settings, including community hospitals and resource-limited regions. This performance-efficiency trade-off, coupled with the demonstrated efficacy of enhancement techniques like RAG and RL, suggests a more accessible implementation pathway where smaller models augmented with these strategies can deliver performance approaching that of much larger models while maintaining computational practicality for real-world clinical deployment.

The complementary benefits of RAG and RL highlight the value of combining memory-based and reward-based approaches in complex optimization tasks. RAG's contribution to final plan quality (11.8% improvement over No-RAG) demonstrates the importance of learning from historical planning attempts—analogous to how experienced human planners build expertise through case exposure. Meanwhile, RL's acceleration of convergence without sacrificing final quality suggests its value in efficiently navigating the optimization landscape. The synergistic integration of these techniques provides a blueprint for future AI systems tackling complex healthcare decision problems that benefit from both experiential knowledge and strategic exploration.

Our study addresses critical gaps associated with both inter-planner variability and the rigid, black-box nature of many existing AI planning approaches. While excessive inter-planner variability can pose risks, variability that arises from thoughtful clinical judgment and patient-specific customization is essential for optimal patient care. Previous research underscores this beneficial variability: planners routinely adjust parameters such as beam angles or dose constraints based on subtle patient factors—adjustments that rigid, fully automated systems might neglect.[2,10] For example, the ability of human planners to recognize anatomical outliers or incorporate patient comorbidities often results in plans that better align with clinical goals than strictly standardized or knowledge-based systems alone.[29,30]

Conversely, existing AI systems, particularly those based on deep learning, often suffer from critical limitations in clinical translation, largely because of their opaque, black-box decision-making processes. The lack of explainability inherent in many deep learning-based solutions makes clinical oversight and validation challenging, hindering trust and adoption among clinicians.[12] Moreover, seemingly optimal plans generated by AI systems may fail practical deliverability criteria due to inherent complexities and hardware limitations.[13,14] For instance, a knowledge-based planning model might produce superior dosimetric outcomes on paper, but the resulting complexity can translate into plans that do not pass standard patient-specific QA measures or strain treatment delivery equipment excessively.

Our LLM-based planning framework uniquely addresses these concerns by combining the benefits of automated planning with human-like reasoning capabilities. The explicit, natural-language reasoning structure used by our LLM agents makes their decision-making process inherently more transparent, potentially overcoming clinician hesitancy due to the opacity of conventional AI methods. Additionally, because our LLM-based agents iteratively optimize based on clinically interpretable metrics and human-understandable prompts, the resulting plans balance optimality with practical deliverability constraints. This explicit reasoning and adaptability represent a fundamental step toward AI systems that enhance, rather than replace, the nuanced clinical judgment necessary for truly personalized radiotherapy treatment.

### 5.3. Privacy-Centric Design and Implementation Considerations

A fundamental innovation of our approach is its privacy-centric design, with the entire system operating within the hospital's secure infrastructure. This local deployment strategy addresses a critical barrier to AI adoption in healthcare—the tension between leveraging advanced AI techniques and protecting sensitive patient information. By eliminating the need for external data sharing, our framework maintains compliance with regulations like HIPAA[7] and GDPR[8] while delivering performance comparable to cloud-based alternatives.

The computational requirements for deploying large language models locally warrant consideration, particularly for the larger parameter (70B+) models. While modern healthcare institutions increasingly possess advanced computational infrastructure, the resource demands may present implementation challenges in some settings. However, several factors mitigate these concerns: (1) the continuing advancement of hardware efficiency for inference, (2) emerging techniques for model compression[47–49] with minimal performance degradation, and (3) the demonstrated efficacy in this work of even the smaller model for generating clinically acceptable plans. Furthermore, the one-time capital investment in computational infrastructure may be justified by the ongoing efficiency gains in planning workflows and the elimination of privacy-related risks associated with external processing.

### 5.4. Limitations and Future Directions

Despite promising results, our study has several limitations that should inform future research. First, the retrospective cohort of 18 prostate cancer patients, while sufficient for proof-of-concept, represents a relatively homogeneous treatment scenario. Given the variability in patient anatomy and physician contouring, a larger cohort is necessary to glean robust statistical conclusions. In this work, performance improvements are presented instead of formal statistics such as p-values, confidence intervals, and effect sizes. Expansion to diverse cancer sites with varying complexity (e.g., head and neck, lung, or central nervous system) would better establish the generalizability of our approach. Second, while our system successfully optimized priority numbers in the existing planning framework, future iterations could control additional parameters such as beam arrangements, collimator angles, and even structure delineation, moving toward more comprehensive planning automation. Third, wide confidence intervals seen in particular cases (such as RAG+RL in Figure 4) suggest potential overfitting due to our limited dataset.

The occasional constraint violations observed (Figure 5) suggest opportunities for refinement in the reinforcement learning reward function. Incorporating adaptive penalties that increase with the magnitude and frequency of constraint violations could further improve plan consistency. Additionally, implementing a clinician-in-the-loop feedback mechanism could allow for mid-process corrections and alignment with individual clinical preferences.[50]

The optimal temperature (T=0.4) was chosen based on a heat map analysis; however this study did not address whether this setting is generalizable to other treatment sites or prescriptions. Additionally, sensitivity of plan score to temperature remains to be quantified and statistically verified in a larger patient cohort. In the absence of a cross-validation study or sensitivity analysis, the temperature selection may be susceptible to overfitting.

From an implementation perspective, future work should explore methods to reduce the computational resources required, particularly for the larger 70B model. In this study, the additional resource and computation use required were not quantified. Additionally, inference latency should be examined in future works to ascertain suitability for clinical settings. Techniques such as model distillation, quantization, and pruning could potentially preserve performance while improving efficiency.[49] Prospective validation in a clinical setting represents the most important next step for this research. Future work must also include an external test or validation dataset from a different institution to understand DOLA's real-world performance. Comparative studies against manually created plans, with evaluation by radiation oncologists, would establish real-world efficacy and acceptability. Such studies should assess not only dosimetric outcomes but also planning efficiency, consistency across planners, and applicability across diverse patient anatomies. Long-term studies could further evaluate whether the improved consistency of automated planning translates to better clinical outcomes through more reliable delivery of optimal dose distributions.

## 6. Conclusion

This autonomous radiotherapy planning agent represents a transformative approach to a critical clinical workflow, addressing inefficiencies, variability, and privacy concerns in traditional planning methods. By integrating LLMs with RAG and RL within a privacy-centric architecture, our system delivers high-quality, personalized treatment plans that meet clinical standards. The demonstrated benefits of larger models and advanced optimization techniques highlight pathways for continued improvement, while the privacy-centric design establishes a model for responsible AI deployment in healthcare. As radiotherapy continues to advance in precision and complexity, automated planning agents like ours will play an increasingly vital role in ensuring that technical capabilities translate to clinical benefits. By reducing planning time, improving consistency, and preserving privacy, this approach has the potential to enhance both operational efficiency and treatment quality—ultimately improving outcomes for cancer patients while establishing a foundation for ethical AI implementation in precision oncology.